\RequirePackage[2020-02-02]{latexrelease}

\documentclass[english,aps,preprint]{revtex4}%
\usepackage[T1]{fontenc}
\usepackage[latin9]{inputenc}
\usepackage{amsmath}
\usepackage{amssymb}
\usepackage{babel}
\usepackage{amsfonts}
\usepackage{graphicx}%
\usepackage{xcolor}
\usepackage{lineno}

\begin{document}
	\title{ From observer-dependent facts to frame-dependent measurement records in
		Wigner friend scenarios}
	\author{J. Allam}
	\author{A. Matzkin}
	\affiliation{Laboratoire de Physique Th\'eorique et Mod\'elisation, CNRS Unit\'e 8089, CY Cergy
		Paris Universit\'e, 95302 Cergy-Pontoise cedex, France}

\begin{abstract}
	The description of Wigner-friend scenarios -- in which external agents
	describe a closed laboratory containing a friend making a measurement --
	remains problematic due to the ambiguous nature of quantum measurements. One
	option is to endorse assumptions leading to observer-dependent facts, given
	that the friend's measurement outcome is not defined from the point of view of
	the external observers. We introduce in this work a model in a relativistic context
	showing that these
	assumptions can also lead to measurement records that depend on the inertial reference
	frame in which the agents make their observations. Our model is based on an
	entangled pair shared by the friend and a distant agent performing space-like
	separated measurements. An external observer at rest relative to the closed
	laboratory and observers in a moving frame do not agree on the observed
	records, which are not Lorentz transforms of one another.

\end{abstract}
\maketitle


\bigskip

\section*{Introduction}

Accounting for measurements remains one of the most intricate theoretical
problems of quantum mechanics. In some instances the theory becomes ambiguous,
and unsurprisingly there is no consensus on how to deal with this ambiguity.
The main difficulty is how to update the quantum state from a premeasurement
linear superposition to a single term upon performing a measurement.

A well-known instance displaying this ambiguity is the Wigner friend scenario
\cite{wigner} (WFS), that has lately seen a renewed interest (see the review
\cite{review} and Refs therein as well as
\cite{lombardi,LP,relano,WFS-WM,toys,histories,rovelli,moreno,samuel,sarkar,wfr}%
).\ Originally introduced as a thought experiment, it has recently been argued
that a universal quantum computer might give rise to a feasible experiment
\cite{rieffel}. The WFS is defined by a (super-)observer W measuring an
isolated laboratory in which an agent F (the friend) makes a measurement on a
quantum system. The ambivalence is whether the super-observer should update
his state upon the friend's measurement, or consider the isolated laboratory
as a quantum superposition evolving unitarily.

Many recent works
\cite{brukner,wiseman,relano,rovelli,samuel,sarkar,wfr,rieffel} embrace a
unitary perspective for W's description.\ Although these works, undertaken
within different interpretative frameworks, take somewhat different views on
quantum measurements, they all assume that the quantum state should only be
updated when an agent observes a measurement record. So in a Wigner friend
scenario F measures the system and updates her state (describing the measured
system and the apparatus), but for W the entire isolated laboratory remains a
closed quantum system and hence evolves unitarily. Facts then become
observer-dependent, as F and W's account and records of the quantum system
measured by F are different.

Another instance in which updating a quantum state after a measurement is
ambiguous happens when relativistic considerations are taken into account. For
example if Alice and Bob measure an entangled pair at space-like separated
locations, in one reference frame A measures first and the state is updated
before B measures, while there are other inertial frames in which the time
ordering of the measurements is inverted and the state is updated and
disentangled before A measures her particle. The current consensus
\cite{peres-review,fayngold} is that state update can be assumed to take place
instantaneously in any reference frame. Having different quantum state
assignments in different frames is inconsequential -- wavefunctions do not
transform covariantly when measurements occur \cite{AA2928}; what only matters
is that the measurement outcomes and their probabilities remain Lorentz
transforms of one another, irrespective of the choice of hypersurface on which
the state is updated.

In this work we will show that introducing relativistic considerations in
unitary accounts of WFS leads to frame-dependent outcomes. This will be done
on the basis of a specific model. \ We will first introduce a probabilistic scenario in which W
and W$^{\prime}$, external observers in two reference frames $\mathcal{R}$ and
$\mathcal{R}^{\prime}$, perform weak measurements \cite{AAV,WVQP} on qubits
correlated with the state of the laboratory. The average position of the
pointers is different in $\mathcal{R}$ and $\mathcal{R}^{\prime}$, leading W
and W$^{\prime}$ to disagree on their observations and also on their
assignments concerning the state of the laboratory. We will then describe more
briefly a slightly different scheme not relying on statistical averages. In
both cases we will require measurement outcomes to be grounded on the
existence of physical records \cite{EPL}.
\textcolor{black}{We will finally discuss the results.
Let us anticipate here that we do not expect frame dependent facts to be
physically possible, so that our results would rather point to a problem
arising when the two assumptions of unitary evolution for the friend and
instantaneous state update in any reference frame are put together.} Note we will
neglect throughout the relativistic character of the spin and the resulting
frame-dependence of entanglement \cite{dunningham18}, thereby assuming a low
velocity for the moving frame.

\begin{figure}[ptb]
\centering
\includegraphics[scale=0.35]{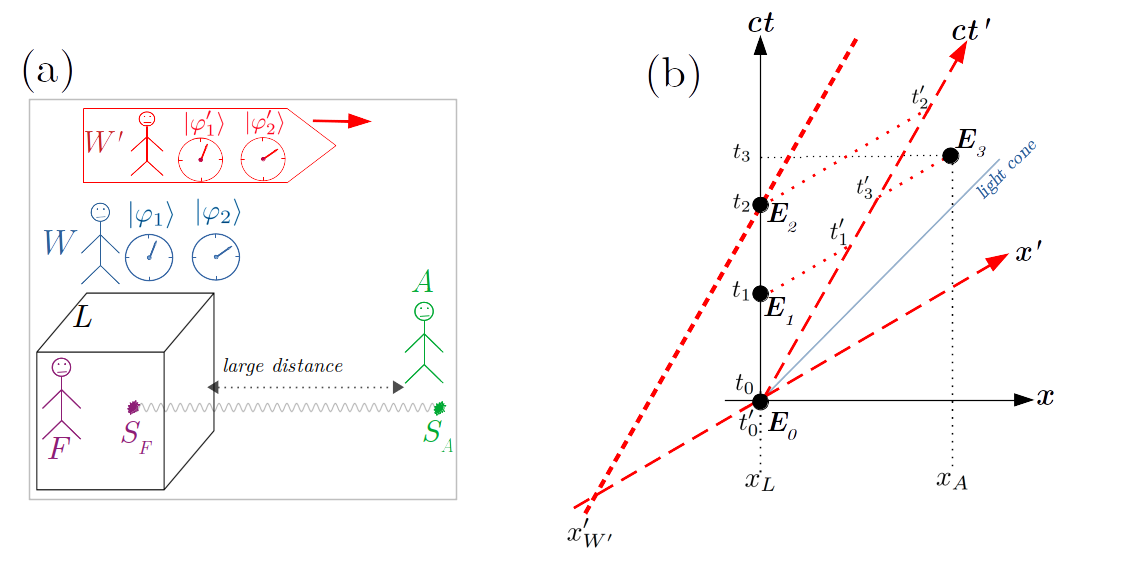} \caption{(a): The scenario involves a
friend F inside a sealed laboratory L, sharing an entangled spin-$\frac{1}{2}$
pair with a distant observer A. F measures her particle S$_{F}$ and then sends
2 qubits outside the lab through a special channel. These qubits are then
measured by an observer W (in $\mathcal{R}$) or W$^{\prime}$ (in
$\mathcal{R}^{\prime}$), either through a weak measurement, or, in another
version of the protocol, by a standard projective measurement. (b): Spacetime
diagram (not to scale) showing events $E_{0}$-$E_{3}$ with respect to
reference frames $\mathcal{R}$ (solid axes ($ct$,$x$)) and $\mathcal{R}%
^{\prime}$ (dashed axes($ct^{\prime}$, $x^{\prime}$)). The time of event
$E_{i}$ in each reference frame is shown along with the space-like
hypersurface in that reference frame (dotted lines). The positions of $x_{L}$
and $x_{A}$ of the lab L and agent A respectively are also shown, as well as
the position of W$^{\prime}$, fixed in $\mathcal{R}^{\prime}$.}%
\label{fig-spacetime}%
\end{figure}

\section*{A Wigner friend protocol}

In our scenario, we consider a friend F in a sealed laboratory L, a
super-oberver W sitting next to L, and a distant observer A (for Alice), at
rest in $\mathcal{R}$ (see Fig. \ref{fig-spacetime}(a)). In this reference
frame, L is positioned at $x=0$, W sits next to L, and A is at $x=x_{A}$, see
Fig. \ref{fig-spacetime}(b). A and F share an entangled state, say $\left\vert
\psi\right\rangle =\alpha\left\vert +\right\rangle _{F}\left\vert
+\right\rangle _{A}+\beta\left\vert -\right\rangle _{F}\left\vert
-\right\rangle _{A}$ where $\sigma_{z}\left\vert \pm\right\rangle
=\pm\left\vert \pm\right\rangle $ (we set $\hbar=1$). We will set
$\alpha=\beta=1/\sqrt{2}$ for simplicity. Inside the friend's laboratory, we
will discriminate the particle whose spin is to be measured denoted
$\left\vert \pm\right\rangle _{F}$, the measurement apparatus with pointer
states $\left\vert m_{k}\right\rangle ,$ and the other degrees of freedom
collectively represented by the environment states $\left\vert \varepsilon
_{k}\right\rangle $. W will make weak measurements on two qubits and require
for that two pointers initially in state $\left\vert \varphi_{1}\right\rangle
$ and $\left\vert \varphi_{2}\right\rangle $ (the wavefunctions $\varphi
_{i}(X_{i})$ can be taken to be Gaussians). E$_{0}$ is the initial event
corresponding to preparation, at $t=0$ in $\mathcal{R}$ and with the quantum
state given by%
\begin{equation}
\left\vert \Psi(t=0)\right\rangle =\left\vert \psi\right\rangle \left\vert
m_{0}\right\rangle \left\vert \varepsilon_{0}\right\rangle \left\vert
\varphi_{1}\right\rangle \left\vert \varphi_{2}\right\rangle .
\end{equation}
\ 

\begin{figure}[ptb]
\centering
\includegraphics[scale=0.15]{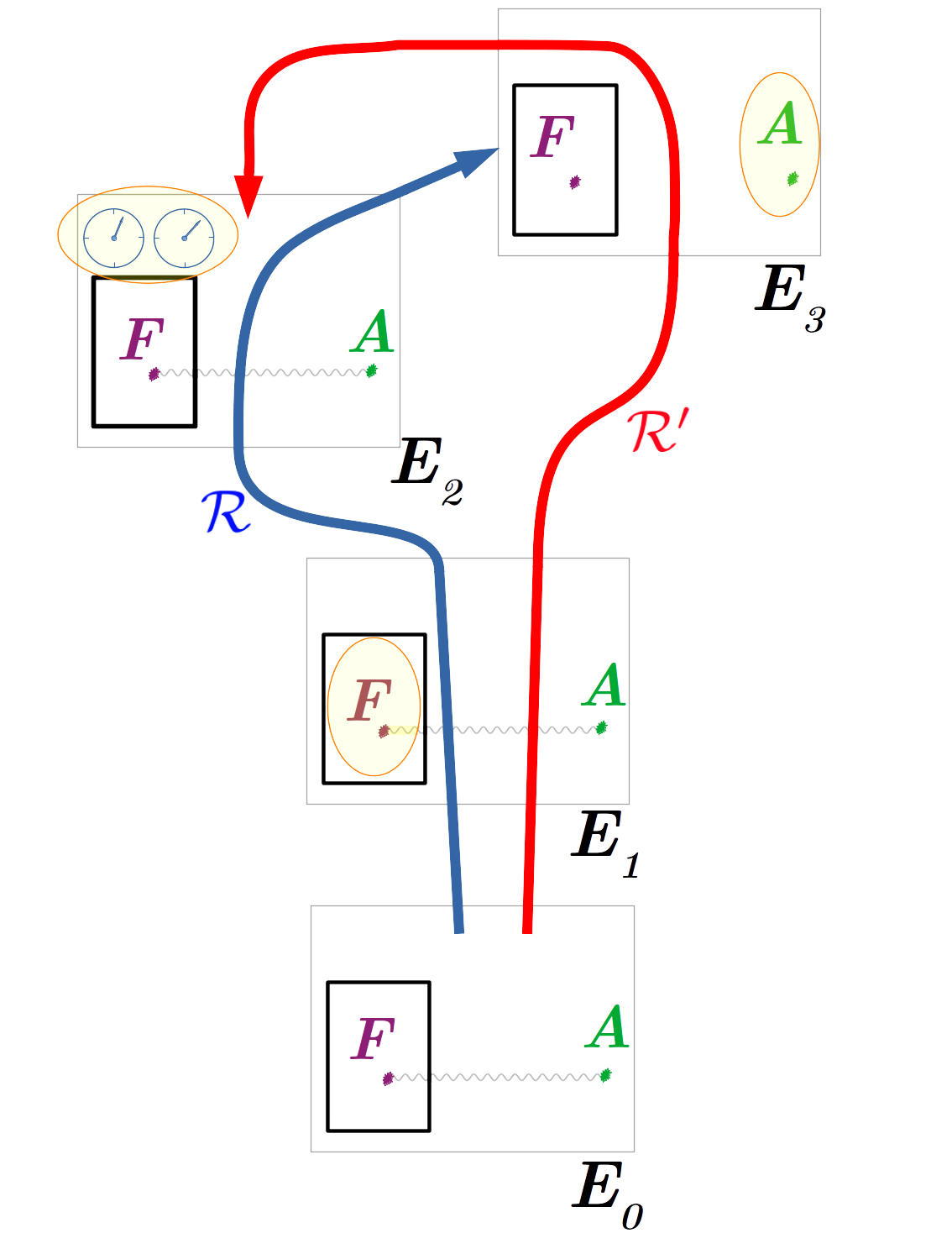} \caption{Time ordering of events
${E_{0},E_{1},E_{2},E_{3}}$ in the two reference frames $\mathcal{R}$
(represented by the blue arrow) and $\mathcal{R}^{\prime}$ (represented by the
red arrow). Event $E_{0}$ represents preparation (see Fig. \ref{fig-spacetime}%
). $E_{1}$ corresponds to $F$ in her sealed lab measuring her qubit and resetting its
state. Event
$E_{2}$ represents F sending 2 qubits outside her lab, where they are
immediately measured by an observer (W or W$^{\prime}$) and E$_{3}$
corresponds to A measuring her qubit. Note that the time ordering of E$_{2}$
and E$_{3}$ is inverted in $\mathcal{R}$ relative to $\mathcal{R^{\prime}}$.
After several runs in which either W or W$^{\prime}$ performs a weak
measurement of the qubits, the average displacement of the pointers weakly
coupled to the qubits can be determined.}%
\label{fig-steps}%
\end{figure}

Event E$_{1}$ corresponds to the friend's measurement: at $t=t_{1}$ the friend
measures the spin component along $z$ by coupling her pointer to the particle
(see Fig. \ref{fig-steps} for the time-ordering of the different steps of the
protocol).\ The quantum state becomes
\begin{equation}
\left\vert \Psi(t_{1})\right\rangle =\alpha\left\vert L_{+}\right\rangle
_{F}\left\vert +\right\rangle _{A}+\beta\left\vert L_{-}\right\rangle
_{F}\left\vert -\right\rangle _{A}\label{t1u}%
\end{equation}
where $\left\vert L_{\pm}\right\rangle \equiv\left\vert \pm\right\rangle
_{F}\left\vert m_{\pm}\right\rangle \left\vert \varepsilon_{\pm}\right\rangle
$ denotes the quantum state of the entire isolated laboratory (we do not
explicitly write the pointer states as they remain unchanged). \textcolor{black}{Then  the friend immediately resets the states of her spin and the
measuring device to a pre-assigned arbitrary state $|s_{0}\rangle|m_{0}%
\rangle.$ Defining $\left\vert \tilde{L}_{\pm}\right\rangle \equiv
|s_{0}\rangle|m_{0}\rangle|\varepsilon_{\pm}\rangle,$ the quantum state thus
becomes%
\begin{equation}
\left\vert \Psi(t_{1})\right\rangle =\alpha\left\vert \tilde{L}_{+}%
\right\rangle _{F}\left\vert +\right\rangle _{A}+\beta\left\vert \tilde{L}%
_{-}\right\rangle_{F} \left\vert -\right\rangle _{A}.\label{t1u1}%
\end{equation}
}Recall that there is no state update despite F's measurement, since Eqs.
(\ref{t1u}) and (\ref{t1u1}) represent the evolution of the entire laboratory for external
observers, including W and A. The friend can open a channel and communicate to
W the fact that she obtained a measurement outcome, eg with an ancilla qubit
\cite{deutsch,sarkar} initially in state $\left\vert 0\right\rangle $ and
becoming $\left\vert 1\right\rangle $ after she completed her measurement, as
long as the state of the ancilla does not get entangled with L (in particular,
the outcome obtained cannot be communicated).

We label by E$_{2}$ the following event: At time $t_{2}$ , F creates 2
particles sent to W by opening a communication channel. The particles sent to
W are qubits prepared in a state identical to the one observed by F after her
previous measurement at $t=t_{1}\ $. \textcolor{black}{Note that for her observation F, a macroscopic
	agent, relies on
the environment states $\left\vert \varepsilon_{\pm}\right\rangle $. }\ The
state after the qubits are created \footnote{Formally we can introduce a vacuum
state $\left\vert 0\right\rangle $ in Eq. (\ref{t1u}) and consider particle
creation operators at the point $r$, $\Phi_{\pm}^{\dagger}(r)=\int
dka^{\dagger}(k)e^{ikr}c(k)$ acting on $\left\vert 0\right\rangle $.
$\Phi_{\pm}^{\dagger}(r)$ creates a particle wavepacket with the spatial
profile determined by $c(k)$ and spin $\pm$.} and sent to W become
\begin{equation}
\left\vert \Psi(t_{2})\right\rangle =\left(  \alpha\left\vert \tilde{L}%
_{+}\right\rangle _{F}\left\vert +\right\rangle \left\vert +\right\rangle
\left\vert +\right\rangle _{A}+\beta\left\vert \tilde{L}_{-}\right\rangle
_{F}\left\vert -\right\rangle \left\vert -\right\rangle \left\vert
-\right\rangle _{A}\right)  \left\vert \varphi_{1}\right\rangle \left\vert
\varphi_{2}\right\rangle .\label{t2u}%
\end{equation}
\newline W immediately makes a weak measurement of the two qubits: first a
spin observable $O$ is coupled to the momentum $P_{i}$ \ of each pointer $i$,
\ i.e. \ we apply \ the \ unitary $\exp\left(  -igOP_{1}\right)  \exp\left(
-igOP_{2}\right)  $ to $\left\vert \Psi(t_{2})\right\rangle $ where $g$ is the
coupling constant between each qubit and its pointer. Then each qubit is
post-selected by making a projective measurement of $\sigma_{\theta_{1}}$ and
$\sigma_{\theta_{2}}$ on qubits 1 and 2 respectively and filtering the
positive outcome in each case (we define $\left\vert +\theta_{i}\right\rangle
=\cos\frac{\theta_{i}}{2}\left\vert +\right\rangle +\sin\frac{\theta_{i}}%
{2}\left\vert -\right\rangle $ and $\sigma_{\theta_{i}}=\overrightarrow
{\sigma}\cdot\overrightarrow{n}_{i}$ where $\overrightarrow{n}_{i}$ makes an
angle $\theta_{i}$ with the $z$ axis and lies in the $xz$ plane). When $g$ is
small (weak coupling) the unitaries can be expanded to first order leading to
an expression that is usually formalized \cite{AAV,WVQP}
as\begin{linenomath}
\begin{align}
\left\langle +\theta_{1}\right\vert \left\langle +\theta_{2}\right\vert
\left.  \Psi(t_{2})\right\rangle  &  =\alpha\left\vert \tilde{L}_{+}\right\rangle
_{F}\left\vert +\right\rangle _{A}\left\langle +\theta_{1}\right\vert \left.
+\right\rangle \left\langle +\theta_{2}\right\vert \left.  +\right\rangle
\left\vert \varphi_{1}^{+}\right\rangle \left\vert \varphi_{2}^{+}%
\right\rangle \label{postR}\\
&  +\beta\left\vert \tilde{L}_{-}\right\rangle _{F}\left\vert -\right\rangle
_{A}\left\langle +\theta_{1}\right\vert \left.  -\right\rangle \left\langle
+\theta_{2}\right\vert \left.  -\right\rangle \left\vert \varphi_{1}%
^{-}\right\rangle \left\vert \varphi_{2}^{-}\right\rangle ,
\end{align}
\end{linenomath}where the pointer states $\left\vert \varphi_{i}^{\pm
}\right\rangle $ are the initial pointer states shifted by the weak values
$\sigma_{i}^{\pm}$, e.g.%
\begin{equation}
\left\vert \varphi_{1}^{-}\right\rangle =e^{-igP_{1}\sigma_{1}^{-}}\left\vert
\varphi_{1}\right\rangle \text{ with }\sigma_{1}^{-}\equiv\frac{\left\langle
+\theta_{1}\right\vert O\left\vert -\right\rangle }{\left\langle +\theta
_{1}\right\vert \left.  -\right\rangle }\label{wvR}%
\end{equation}
and similarly in the other cases. For definitiness we will set $O=\sigma_{z}$,
so that the $z$ spin component is measured weakly for both qubits.

Right after post-selection (we neglect the duration $\tau$ of the weak
measurement \cite{WVQP}) W measures the position $X_{1}$ and $X_{2}$ of each
pointer. By repeating the experiment, he collects statistics (when
post-selection is successful) to determine the average pointer positions.
Theoretically the average is obtained from Eq. (\ref{t2u}) by tracing out the
friend's and Alice's degrees of freedom in order to obtain the reduced density
matrix $\rho_{12}$. Recalling that by construction the average position is
$\int X_{i}\left\vert \varphi_{i}^{\pm}(X_{i})\right\vert ^{2}dX_{i}%
=g\sigma_{i}^{\pm}$, and labeling the post-selection projectors $\Pi
_{+\theta_{i}}\equiv\left\vert +\theta_{i}\right\rangle \left\langle
+\theta_{i}\right\vert $ we obtain for the choices made here
\begin{equation}
\left\langle X_{1}X_{2}\right\rangle =\text{Tr}\left(  \rho_{12}\Pi
_{+\theta_{1}}\Pi_{+\theta_{2}}X_{1}X_{2}\right)  =\frac{g^{2}}{4}\left(
1+\cos\theta_{1}\cos\theta_{2}\right)  . \label{avR}%
\end{equation}
Note that this result is obtained by W\ by observing his own pointers,
irrespective of the outcomes that can be attributed to F. Finally the event
E$_{3}$, at $t_{3}$ in $\mathcal{R}$ corresponds to A measuring her qubit in
the $\left\vert \pm x\right\rangle $ basis ( $\left\vert \pm x\right\rangle
=\left(  \left\vert +\right\rangle \pm\left\vert -\right\rangle \right)
/\sqrt{2}$); she obtains her local predictions for her qubit's outcome by
tracing out all the other degrees of freedom.

Let us now consider the same protocol but for observers in a different
inertial frame $\mathcal{R}^{\prime}$ (see Fig.\ \ref{fig-spacetime}).
$\mathcal{R}^{\prime}$ is chosen such that E$_{3}$ happens before E$_{2}$,
i.e. A measures her spin before F sends her qubits outside the lab. The time
ordering is $t_{0}^{\prime}<t_{1}^{\prime}<t_{3}^{\prime}<t_{2}^{\prime}$ (see
Fig. \ref{fig-steps}). \textcolor{black}{F makes her measurement first followed by spin reset, as per Eqs. (\ref{t1u})-(\ref{t1u1}). But after A's measurement
$\left\vert \Psi^{\prime}(t_{1}^{^{\prime}})\right\rangle =\alpha\left\vert
\tilde{L}_{+}\right\rangle _{F}\left\vert +\right\rangle _{A}+\beta\left\vert
\tilde{L}_{-}\right\rangle _{F}\left\vert -\right\rangle _{A}$ is updated to
either
\begin{equation}
\left\vert \Psi_{+x}^{\prime}(t_{3}^{^{\prime}})\right\rangle =\frac{1}%
{\sqrt{2}}|s_{0}\rangle|m_{0}\rangle\big(\left\vert \varepsilon_{+}%
\right\rangle +\left\vert \varepsilon_{-}\right\rangle \big)\equiv
|s_{0}\rangle|m_{0}\rangle\left\vert \varepsilon_{+x}\right\rangle \label{pra}%
\end{equation}
or to
\begin{equation}
\left\vert \Psi_{-x}^{\prime}(t_{3}^{^{\prime}})\right\rangle =\frac{1}%
{\sqrt{2}}|s_{0}\rangle|m_{0}\rangle\big(\left\vert \varepsilon_{+}%
\right\rangle -\left\vert \varepsilon_{-}\right\rangle \big)\equiv
|s_{0}\rangle|m_{0}\rangle\left\vert \varepsilon_{-x}\right\rangle
.\label{prb}%
\end{equation}
We have used the notation $|\varepsilon_{\pm x}\rangle\equiv\big(\left\vert
\varepsilon_{+}\right\rangle \pm\left\vert \varepsilon_{-}\right\rangle
\big)/\sqrt{2}$ to emphasize that
the superposition of the environment states $\left\vert \varepsilon
_{+}\right\rangle \pm\left\vert \varepsilon_{-}\right\rangle $ can be taken to 
represent the
state of the environment that would be obtained had $\sigma_{x}$ been
measured \footnote{This can be seen
	for example if the friend measures $\sigma_{z}$ on a spin in state $\left\vert
	+x\right\rangle $, resets the pointer to $\left\vert m_{0}\right\rangle ,$
	and measures $\sigma_{x}$. For an external observer the lab is left in
	a superposition of states $\left\vert \pm x\right\rangle \left\vert m_{0}\right\rangle \left(
	\left\langle \pm  x\right\vert \left.  + \right\rangle \left\vert \varepsilon
	_{+}\right\rangle +\left\langle \pm x\right\vert \left.  -\right\rangle
	\left\vert \varepsilon_{-}\right\rangle \right)  $. If F deletes all records 
	of the previous $\sigma_{z}$ measurement, the term between
	$\left(  ...\right)  $ must correspond to $\left\vert \varepsilon
	_{\pm x}\right\rangle $. }.\ Note however that in order to run our
argument we just need $\left\vert \varepsilon_{\pm x}\right\rangle $ to be
macroscopically different from $\left\vert \varepsilon\pm\right\rangle $.
Indeed the friend is also a macroscopic object and any observation of a
quantum result is distilled by the environment \cite{zurek}. Consequently, the
superpositions (\ref{pra})-(\ref{prb}) resulting in either $\left\vert
\varepsilon_{+x}\right\rangle $ or $\left\vert \varepsilon_{-x}\right\rangle $
correspond to a state in which the friend now observes a spin outcome
different from the original measurement basis.} The records of the original
measurement of $\sigma_{z},$ \textcolor{black}{encapsulated in the environment states
$\left\vert \varepsilon_{\pm}\right\rangle $} have been erased \cite{wiseman,EPL} by the
superpositions, \textcolor{black}{(\ref{pra})-(\ref{prb}), which can be rewritten as
\begin{equation}
\left\vert \Psi_{\pm x}^{\prime}(t_{3}^{^{\prime}})\right\rangle =\left\vert
\tilde{L}_{\pm x}\right\rangle _{}
	.\label{prc}%
\end{equation}
}

Following our protocol, E$_{2}$ takes place at time $t_{2}^{\prime}$: F opens
a channel and sends the 2 qubits outside L to W$^{\prime}$, an observer at
rest in $\mathcal{R}^{\prime}$ and moving with respect to F. In $\mathcal{R}%
^{\prime}$ the state updates (\ref{pra}) and (\ref{prb}) imply that the qubits
sent by F will be in accordance with her observation at $t_{3}^{\prime}$,
\textcolor{black}{namely $\left\vert +x\right\rangle $ if the environment is in state
$\vert\varepsilon_{+x}\rangle$ or $\left\vert -x\right\rangle $ if the
environment is in state $\vert\varepsilon_{-x}\rangle$.} The resulting quantum
state at the time W$^{\prime}$ measures the qubits is  \textcolor{black}{
\begin{equation}
\left\vert \Psi_{+x}^{\prime}(t_{2}^{^{\prime}})\right\rangle 
 =\left\vert
\tilde{L}_{+x}\right\rangle  \left\vert +x\right\rangle \left\vert +x\right\rangle
\label{Rp1}%
\end{equation}
or
\begin{equation}
\left\vert \Psi_{-x}^{\prime}(t_{2}^{^{\prime}})\right\rangle  =\left\vert
\tilde{L}_{- x}\right\rangle  \left\vert -x\right\rangle \left\vert -x\right\rangle .
\label{Rp2}%
\end{equation}
} 
After the weak measurements of
the 2 qubits by W$^{\prime}$, an observer in $\mathcal{R}^{\prime}$ will
describe the shifted pointer states as either $\left\vert \varphi_{1}^{\prime+
x}\right\rangle \left\vert \varphi_{2}^{\prime+ x}\right\rangle $ or
$\left\vert \varphi_{1}^{\prime- x}\right\rangle \left\vert \varphi
_{2}^{\prime- x}\right\rangle $
where for example%
\begin{equation}
\left\vert \varphi_{1}^{\prime-x}\right\rangle =e^{-igP^{^{\prime}}_{1}%
\sigma_{1}^{-x}}\left\vert \varphi_{1}^{\prime}\right\rangle \text{ with
}\sigma_{1}^{-x}\equiv\frac{\left\langle +\theta_{1}\right\vert O\left\vert
-x\right\rangle }{\left\langle +\theta_{1}\right\vert \left.  -x\right\rangle
} \label{wvRp}%
\end{equation}
and similarly for the other pointer states. Note that the pointer shifts in
$\mathcal{R}$ and $\mathcal{R}^{\prime}$ are different [compare Eqs.
(\ref{wvR}) and (\ref{wvRp})]. The average shifts will also be different; from
the mixed density matrix, and recalling that we have taken $O=\sigma_{z}$ (and
that we have assumed the spin is not affected by Lorentz transformations), we
obtain%
\begin{equation}
\left\langle X_{1}^{\prime}X_{2}^{\prime}\right\rangle =g^{2} \cos\theta
_{1}\cos\theta_{2}. \label{avRp}%
\end{equation}
This is different than the average observed in $\mathcal{R}$. The difference does not
depend on any parameters characterizing the Lorentz transformation between the
two frames \textcolor{black}{but on the fact that F sends qubits that are different in each frame.}
 Having frame-dependents averages can hardly lead to a
consistent picture, as will perhaps be clear by slightly modifying our
scenario in order to obtain the frame-dependent character without relying on statistics.


\section*{A modified scenario with projective measurements}

In the modified protocol, W (or W$^{\prime}$) does not perform a weak
measurement on each qubit sent by F, but makes a standard projective
measurement on each qubit, one in the $\left\vert \pm\right\rangle $ basis,
the other in the $\left\vert \pm x\right\rangle $ basis. In $\mathcal{R}$, we
see from Eq.\ (\ref{t2u}) that \textcolor{black}{this breaks the superposition of lab states $\left\vert \tilde{L}_{\pm}\right\rangle $ and leads to a definite state for F's
environment, either $\left\vert \varepsilon_{+}\right\rangle $ or $\left\vert
\varepsilon_{-}\right\rangle $ in which the physical record is embedded. F can declare the measurement outcome obtained at $t_{1}$, $+$ or $-$, for example by sending a huge number of qubits in states $\left\vert +\right\rangle $ or  $\left\vert -\right\rangle $ that W can analyze.} Hence, F's
outcome that was defined solely for F in the time interval $t_{1}<t<t_{2}$
becomes public and objective for observers outside the lab at $t>t_{2}$.
Again, Alice measures her qubit at $t_{3}$ but this does not change F's
record, since the lab was opened and the state update took place at $t_{2}$.

In $\mathcal{R}^{\prime}$, A's measurement takes place before F sends her
qubits, so that W$^{\prime}$'s projective measurements on the qubits take
place after the state update.\ From Eqs. (\ref{Rp1})-(\ref{Rp2}), we see that
W$^{\prime}$'s qubits measurements leave again the lab environment in a
definite state but this time for an observer in $\mathcal{R}^{\prime}$ the
assigned state is \textcolor{black}{ $\left\vert
\tilde{L}_{+ x}\right\rangle  $ or $\left\vert
\tilde{L}_{- x}\right\rangle  $}. At this stage F can declare her outcome in a similar way.

This modified protocol therefore leads to different physical records being
observed in $\mathcal{R}$ and $\mathcal{R}^{\prime}$, \textcolor{black}{which as mentioned
above is a consequence of the unitary quantum description of an agent making
interventions that precisely depend on the quantum state at intermediate times, a state that is different 
 in the two reference frames.} Note that we now rely on
the friend's outcomes, assuming the laboratory can be opened unambiguously if
it is in a definite eigenstate, at which stage F's record becomes objective.
In the statistical version of the protocol the frame-dependent outcomes were
the external weakly-coupled pointers.\ In neither case do we assume that the
laboratory is directly measured by W or W$^{\prime}$.

\section*{Discussion and Conclusion}

While dealing with observer-dependent facts might be considered to be a viable
option (it is consistent since the observers are isolated from one another and
the alternative facts cannot be defined jointly \cite{brukner}), having facts
depending on an observer's reference frame cannot be accepted; for example,
for the present protocol it can lead to signalling \cite{cloning}. Evidently,
relaxing the assumption that the laboratory can be described by a quantum
state or, if this is deemed possible, relaxing the hypothesis that (for
practical or fundamental reasons) its evolution can be described unitarily
avoids the inconsistency.\ Note that in the latter case, this means that a
measurement by an observer leads all the other agents to update their state
assignments (to a mixed density matrix). In this case, it can be readily
verified that\ although the quantum states at intermediate times will be
different in $\mathcal{R}$ and $\mathcal{R}^{\prime}$, the observers in all
reference frames will agree on the outcomes and probabilities of each event
(in general the density matrices are then connected by a Lorentz transformation).

Notwithstanding, it is worthwhile to examine the possibility that more
specific assumptions could play a role. First, while an isolated macroscopic
system might well be accounted for unitarily, demanding an agent's arbitrary
operations to be described with unitaries implies stronger constraints. Indeed
it has been argued that a valid measurement (as opposed to the creation of
correlations) cannot rely on a global unitary evolution -- it requires an
``intervention'' \cite{peres-classical}. Hence certain operations realized by
an agent might not be possible to model by assuming a perfect correlation
between the state of the measured spin and the quantum state of the
laboratory. Second, when dealing with entangled states, we have treated the
laboratory, a very massive complex system, the same way we would account for
entangled elementary particles, such as photons or electrons. This might not
be correct, either for practical reasons due to decoherence \cite{fuentes}, or
for fundamental reasons, e.g. if, as advocated by Bell \cite{bell1,bell2},
there is a preferred quantum frame \cite{preferred} that would change the
state update rule or the entanglement properties in a mass-dependent way.

To sum up, we have introduced a model for a Wigner friend scenario for which
assuming the friend and Wigner observe different facts also leads to observers
in different reference frames disagreeing on their observations. In a
statistical version of the model, an external observer in one or the other
reference frame performs weak measurements on qubits correlated with the state
of the laboratory and the average state of the qubit pointers was seen to
depend on the reference frame. In a one-shot version of the model the external
observer indirectly opens the laboratory by making projective measurements on
the qubits, and this leads the friend to declare outcomes that are different
in each reference frame.\ We briefly discussed which assumptions could be
relaxed in order to avoid frame-dependent results. More work is needed in
order to understand how to account quantum mechanically for physical
operations carried out by agents, especially in view of future experimental
realizations of such scenarios with quantum computers.

\end{document}